\documentclass[a4paper,11pt]{article}
\usepackage{jheppub} % for details on the use of the package, please
\usepackage[T1]{fontenc} % if needed
 \usepackage{bm}% bold math
 \usepackage{ulem}
\newcommand{\Ds}{\displaystyle}                           %%%%%%%%%
\newcommand{\be}{\begin{equation}}\newcommand{\ee}{\end{equation}}%
\newcommand{\ba}{\begin{eqnarray}}                        %%%%%%%%%
 \newcommand{\ea}{\end{eqnarray}}                         %%%%%%%%%
\def\MSbar{\relax\ifmmode\overline                        %%%%%%%%%
           {\rm MS}\else{$\overline{\rm MS}${ }}\fi}     %%%%%%%%%
   \def\eg{\hbox{\it e.g.}{ }} %\def\cf{\hbox{\it cf.}{ }} %%%%%%%%

\newcommand{\dAA}{d_A^{abcd}d_A^{abcd}}
\newcommand{\dRR}{d_F^{abcd}d_F^{abcd}}
\newcommand{\dRA}{d_F^{abcd}d_A^{abcd}}                                           %%%%%%%%%
\newcommand{\dRRNA}{\frac{d_F^{abcd}d_F^{abcd}}{N_A}}
\newcommand{\dRANA}{\frac{d_F^{abcd}d_A^{abcd}}{N_A}}
\newcommand{\dAANA}{\frac{d_A^{abcd}d_A^{abcd}}{N_A}}

\usepackage{color}
\definecolor{green}{rgb}{0.133,0.56,0}
\definecolor{DarkGreen}{rgb}{0.04,0.5,0.1}

   \newcommand{\RedTn}[1]{\textcolor{red}{#1}}
\def\1{\hbox{{1}\kern-.25em\hbox{l}}}% Web of Conferences font
\title{
Adler function, Bjorken polarized Sum Rule: confirmation of elements of the $\{\beta\}$-expansion\\ and the diagrams.
}
  \author[a]{ S.~V.~Mikhailov}
   \affiliation[a]{Bogoliubov Laboratory of Theoretical Physics, JINR,
                141980 Dubna, Russia}
\emailAdd{mikhs@theor.jinr.ru}
\abstract{
Different ways exist to obtain the elements of the $\{\beta \}$-expansion for renormgroup invariant quantities.
Here  we consider independent confirmation within the standard QCD of a number of our results \cite{Baikov:2022zvq} for the values of elements of
 this expansion for the nonsinglet Adler $D_A$-function,  Bjorken polarized sum rules $S^{Bjp}$ up to the order  N$^4$LO.
 We suggest an approach to estimate the results of high order QCD calculations using a smaller number of diagrams of the specific type.
This type is based on a proposed generalization of  Naive NonAbelianization.
}

\begin{document}
\maketitle

\section{Introduction}
\label{intro}
Here we will consider the results of $\{\beta \}$--expansion \cite{Mikhailov:2004iq} and discuss the reliability of the obtained elements of this expansion \cite{Baikov:2022zvq} and the relation of the latter with the corresponding Feynman diagrams.
The elements of this expansion appear at the decomposition of the standard perturbative coefficients
of any renormalization group invariants (RGIs);
here these coefficients are $d_n$ or $c_n$  for the nonsinglet Adler $D_A$-function, or the Bjorken polarized sum rule (BpSR) $S^{Bjp}$  respectively.
We consider independent  confirmation of our results for the decomposition of the nonsinglet Adler $D_A$-function for
the values of elements $d_{3}[.]$ ($c_{3}[.]$) of $\{\beta \}$--expansion
and  for some elements of $d_{4}[.]$ \cite{Baikov:2022zvq}.
The agreements are obtained using the already known results of N$^{3,4}$LO  calculations \cite{Ball:1995ni} of some diagrams in
the \MSbar-scheme in pure QCD.
In  the case of the nonsinglet $D_A$-function at the default choice of  scale $Q^2=\mu^2$,
\vspace*{-2mm}
\begin{eqnarray} \label{eq:PT-D-C}
  D_{A}(a(\mu^2)) = 1 + \sum_{n\geq1} a^n_s(\mu^2)~d_n, ~~a_s(\mu^2)\!=\!\alpha_s(\mu^2)/(4\pi),
 \end{eqnarray}
the decomposition of $d_n$ looks like \cite{Mikhailov:2004iq,Baikov:2022zvq},\vspace*{-2mm}
\begin{subequations}
\label{eq:d_beta}
\begin{eqnarray} \vspace*{-2mm}
a_s^2~~d_2&=&\! \beta_0\,d_2[1]
  + d_2[0]\, ,\label{eq:d_2}\\
a_s^3~~  d_3&=&\!
  \beta_0^2\,d_3[2]
  + \beta_1\,d_3[0,1]
  +       \beta_0 \,  d_3[1]
  + d_3[0]\, ,\label{eq:d_3} \\
a_s^4~~  d_4
   &=&\! \beta_0^3\, d_4[3]
     + \beta_1\,\beta_0\,d_4[1,1]
     + \beta_2\, d_4[0,0,1]
     + \beta_0^2\,d_4[2]
     + \beta_1  d_4[0,1]
     + \beta_0\,d_4[1] \nonumber \\
   && \phantom{\beta_0^3\, d_4[3]+ \beta_1\,\beta_0\,d_4[1,1]+ \beta_0^2\,}~+d_4[0]\,,
       \label{eq:d_4}  \vspace*{-2mm} \\
 \ldots &=& \ldots   \nonumber \\
%\end{eqnarray}
%\begin{eqnarray}
%  \ldots &=& \ldots   \nonumber \\
a_s^n~~ d_n&=& \beta_0^{n-1} d_n[n-1]+ \ldots + d_n[0]\,,    %\nonumber
\end{eqnarray}
 \end{subequations}
 where $\beta_i$ are the coefficients of expansion of the QCD $\beta$-function (see Appendix \ref{App:A}).
The $\{\beta \}$-expansion shows how the intrinsic charge renormalization manifests itself;
thus all  possible $\beta_i$-terms  appear in each order $n$ of perturbative expansion (PT).
 This property differ it from more or less formal expansions over powers of selected lagrangian parameters or Casimirs,
  say the number of active quark flavors $n_f$, or powers of $\beta_0$ \cite{Lovett-Turner:1995zwc}, which is based on the same formal reasons.
The procedure of calculating  the expansion elements in (\ref{eq:d_beta}) was based, in its turn, on the results  of extended QCD (QCDe)
 presented in \cite{Chetyrkin:2022fqk}, which possesses a number of new degrees of freedom (d.o.f.).
  These d.o.f. are arbitrary number of different \textit{fermion representations}, revealing themselves only in intrinsic loops.
  After decomposition as in (\ref{eq:d_beta}),
 one can equate the contributions of these new d.o.f.  to zero and return from QCDe to the standard QCD.
  Note here that now are at least two different approaches to obtain the values of the elements within the standard QCD that lead to different results.
One of them, started with \cite{Mojaza:2012mf}, is based on an interpretation of RG transformation, it got the name
``Principle of Maximum Conformality'' (PMC), see review  \cite{DiGiustino:2023jiq} and references therein.
The other approach, see \cite{Cvetic:2016rot},
is based on an extended interpretation of Crewther-Broadhurst-Kataev (CBK) relation \cite{Broadhurst:1993ru}.
So, some doubt appeared about how reliable the results for the elements obtained within QCDe in \cite{Baikov:2022zvq} are with respect to the standard QCD.
% \cite{Kataev:2022iqf,Goriachuk:2021ayq}.

Here we %completely
confirm our results in order $O(a_s^3)$ and confirm them in part in order of $O(a_s^4)$ based on the calculation \textit{within the standard} QCD and focusing on the diagrammatic origins of
the elements of the $\{\beta \}$-expansion.
In addition in Sec.\ref{sec:4} we propose a hierarchy of contributions to the coefficients $d_n (c_n)$ based on a certain generalization of Naive Nonabelianization (NNA) in the framework of the $\{\beta \}$-expansion.
The usage of this hierarchy can facilitate the laborious  process of estimating physical quantities in high-loops calculations.
%\setcounter{equation}{0}

%\newpage
\section{Test for $\{\beta \}$-expansion elements of $D_A\, (S^{Bjp})$ at $O(a_s^3)$  }
 \label{sec:2.2}
Recall the  elements of the $\{\beta \}$-expansion for  $D_A$  up to the order $O(a_s^3)$
\cite{Mikhailov:2004iq,Baikov:2022zvq,Kataev:2014jba}, where the first term $d_1=3\,C_\text{F}$ :
%\footnote{a missprint in the expression
%for $d_3[1]$ in articles \cite{Kataev:2010du,Kataev:2014jba} had been corrected in \cite{Mikhailov:2016feh} }
 \begin{eqnarray}
d_2[1]&=&d_1\left(\frac{11}2-4\zeta_3\right);~
d_2[0]=d_1\left(\frac{\rm C_A}3-\frac{\rm C_F}2\right); \label{D-21}
\end{eqnarray}
\begin{subequations}
\label{eq:d1-4}
 \begin{eqnarray}
d_3[2]&=&d_1\left(\frac{302}9-\frac{76}3\zeta_3\right);~d_3[0,1]=d_1\left(\frac{101}{12}-8\zeta_3\right);\label{D-32}\\
d_3[1]&=&
    d_1\left[{\rm C_A}\left(-\frac{3}4 + \frac{80}3\zeta_3 -\frac{40}3\zeta_5\right) -
    {\rm C_F}\left(18 + 52\zeta_3 - 80\zeta_5\right) \label{D-31}\right]; \label{eq:d31} \\
d_3[0]&=& d_1\Bigg[{\rm C_A^2}\left(\frac{523}{36}- 72 \zeta_3\right)
    +\frac{71}3 {\rm C_A C_F} - \frac{23}{2} {\rm C_F^2}\Bigg]~.  \label{D-30}
\end{eqnarray}
\end{subequations}
These decompositions for $d_3$ were first obtained in \cite{Mikhailov:2004iq} where one used
the  results for $D_A$ with a single additional degree of freedom -- the light gluino \cite{Chetyrkin:1996ez},
then the elements of this decomposition were
confirmed in \cite{Baikov:2022zvq} within  QCDe with any numbers of fermion d.o.f.
Below we reproduce the same values of the elements $d_3[.]$ by considering the results for specific diagrams
for $D_{A}$ calculated in \textit{pure} QCD.

Let us start with the simplest case in (\ref{D-21}).
The first element $d_2[1]$ originates from the diagram where the single gluon line
is dressed in a quark bubble, which is proportional to $ T_R n_f \sim \beta_0$.
Just this correspondence allows one to perform the decomposition unambiguously.
The sequential dressing of this only one gluon line generates its anomalous dimension that is finally revealed as a set of $\beta$-elements of our interest also in the higher orders of PT.
\begin{figure}[ht]
\centerline{\includegraphics[width=0.2\textwidth]{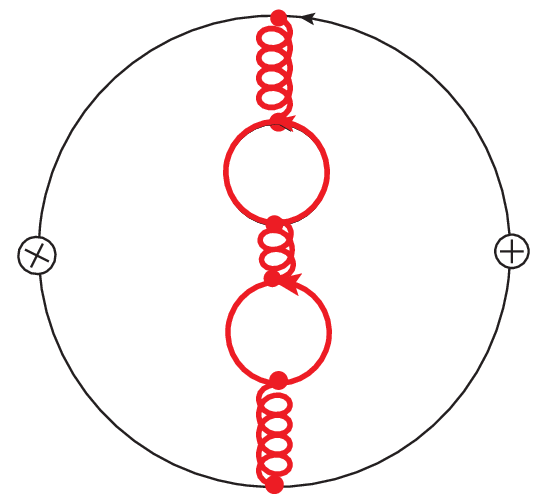}~~~~~~~~~~ %
  \includegraphics[width=0.2\textwidth]{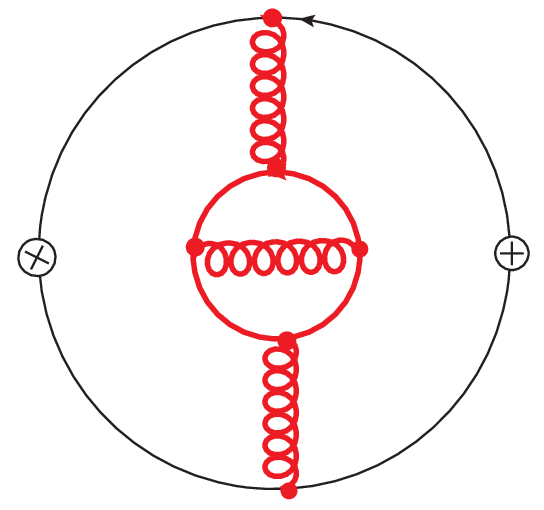}
 }
\vspace{-3mm} \caption{\footnotesize{\label{fig:fig1}
Diagrams for $D_A$ in order $a_s^3$ with self-energy dressing of only one gluon line (red in color).
One should add to these samples of diagrams  all possible connections of gluon lines.
\textbf{Left.} proportional to $C_\text{F}\cdot (T_R n_f\equiv x)^2$, $\sim \beta_0^2$. \textbf{Right.} proportional to
$C^2_\text{F} (T_R n_f\equiv x )$, $\sim \beta_1$} \,.}
\end{figure}
In the next order the 2nd loop dressing of only one gluon line generates  specific
diagrams with two 1-loop bubbles (reducible), see Fig.\ref{fig:fig1}(Left)
and  with 2-loop one (irreducible), Fig.\ref{fig:fig1}(Right).

We reproduced in Fig.\ref{fig:fig1}  the diagrams taken from
Fig.C1(IV1) and Fig.C1(IV2) in \cite{Ball:1995ni},
and the abelian projection of the results corresponding to them was presented in Appendix C, Eqs.(C.2),(C.3) there.
The result for the diagram in Fig.\ref{fig:fig1}(Left) independently confirms the value of the $d_3[2]$ term in (\ref{D-32})
and coincides also with the general expression for $d_n[n-1]$,
which can be extracted from the results of \cite{Broadhurst:1992si,Broadhurst:1993ru}.
The diagram in the right panel contributes to the term $\beta_1 d_3[0,1]$ by means of
anomalous dimension of a gluon in order of $a_s^2$.
The latter, in its turn, includes the unique  Casimir $ C_\text{F} \cdot (T_R n_f\equiv x)$, which is common with $\beta_1$,
see Eq.(\ref{eq:A2}).
This diagrammatic correspondence unambiguously  determines the element $d_3[0,1]$ within the standard QCD calculation
that value coincides with the one in (\ref{D-32}).
Further we will trace the contributions to $d_n$ of the color structures starting with maximum numbers of quark loops:
the evident case of a renormalon chain with only one gluon line contribution gives us $C_F\cdot x^{(n-1)}$,
then  one fermion loop less -- $C_F (C_F x^{(n-2)})$, then $C_F(C_F^2 x^{(n-3)})$ and so on.
In other words, these terms are a projection onto the abelian
sector of the QCD calculation (proportional to QED) of $D_A$ and $S^{Bjp}$.
The abelian part of the structure of $\beta$-coefficients is well recognized  in the first terms
(underlined) of the corresponding
formulas in Appendix \ref{App:A}.

Now we will show that independent knowledge of $d_3[0,1]$ guarantees also independent determination of
two remainder elements $d_3[1], d_3[0]$ of this order if the coefficient $d_3$ as a whole is already known.
Using the definition $T_R n_f=x$, we introduce \cite{Mikhailov:2016feh} the consequence  of roots $x_0,x_1, \ldots \sim x$  of the QCD $\beta$-function coefficients
 $\beta_0(x_0)\!=\!0$, $\beta_1(x_1)\!=\!0, \ldots$, respectively,

\begin{subequations}
\ba
\!\!\!\!\!\!\!\! x_0\!=\!\Ds \frac{11}{4}{\rm C_A};&& \\
\!\!\!\!\!\!\!\!\beta_0(x_0)=0;&\!\beta_1(x_0)\!=\!-{\rm C_A}\!\left(7{\rm C_A}+11{\rm C_F} \right)\!;&\!\!\beta_2(x_0)\!=\!{\rm C_A}\! \!\left(\!\frac{11}{2}{\rm C_F^2-\frac{1127}{24} {\rm C_A^2}\! -\! \frac{77}3{\rm C_A}{\rm C_F}\!}\! \right)\!\!;\label{eq:x_0}\\
\!\!\!\!\!\!\!\!&\Ds x_1\!=\!
\frac{17}{2}\frac{\rm C^2_A}{5{\rm C_A}+3{\rm C_F} };& \\
\!\!\!\!\!\!\!\! \beta_0(x_1)\!=&\!\!\!\Ds {\rm C_A}\frac{7{\rm C_A}+11{\rm C_F}}{5{\rm C_A}+3{\rm C_F}}; \beta_1(x_1)=0;&\beta_2(x_1)=\ldots\,. \label{eq:x_1}
\ea
\end{subequations}
Introducing a reduced $\tilde{d}_3(x)\!\equiv\!d_3(x) - \beta_0^2(x) d_3[2]$, depending on the argument
 $x$, we obtain from the $\{\beta \}$-expansion for $d_3$ (Eq.(1.3) in \cite{Baikov:2022zvq}),
\begin{subequations}
 \label{eq:d_3result}
\begin{eqnarray}
%\beta_0(x_0)=0;~~\beta_1(x_1)=0;&& \\
\tilde{d}_3(x_0)- \beta_1(x_0)d_3[0,1]&=&d_3[0]; \\
\frac{\tilde{d}_3(x_0)-\tilde{d}_3(x_1)}{\beta_0(x_1)}+ \frac{\beta_1(x_0)}{\beta_0(x_1)} d_3[0,1]&=&d_3[1]\,,
\end{eqnarray}
 \end{subequations}
where $d_3(x)$ (and $d_4(x)$) are known from the original calculations \cite{Baikov:2010je}.
So we get \textit{independent confirmation} within QCD of all the results for $d_3[.]$ displayed in Eqs.(\ref{eq:d1-4}).

It is clear that the same relations, like Eq.(\ref{eq:d_3result}), is true also for the elements $c_3[.]$ in the $\{\beta \}$-expansion of $S^{Bjp}$.
The required value of $c_3[0,1]$ can be obtained from $d_3[0,1]$ based on the CBK relation \cite{Kataev:2010du,Baikov:2022zvq},
\ba \label{eq:CBK1}
\!c_2[1]+d_2[1]\!&=&\!c_3[0,1]+d_3[0,1]\!=\!\!c_n[\underbrace{0,0,\ldots, 1}_{n-1}] +
d_n[\underbrace{0,0,\ldots, 1}_{n-1}]\!=\!
3C_\text{F}\! \left(\frac{7}2-4\zeta_3 \right) \label{eq:c4__1+d4__1}.
\ea
Therefore, the knowledge of this $d_3[0,1]$ makes it possible to restore both $d_3[.]$ and $c_3[.]$ only within QCD calculations and confirms
the results obtained with the help of QCDe in \cite{Baikov:2022zvq}, and within QCD with light gluino \cite{Mikhailov:2004iq}.
%\newpage
\section{  Test for $\{\beta \}$-expansion elements of $D_A$ at $O(a_s^4)$}
 \label{sec:sec3}
\begin{figure}[ht]
\centerline{\includegraphics[width=0.17\textwidth]{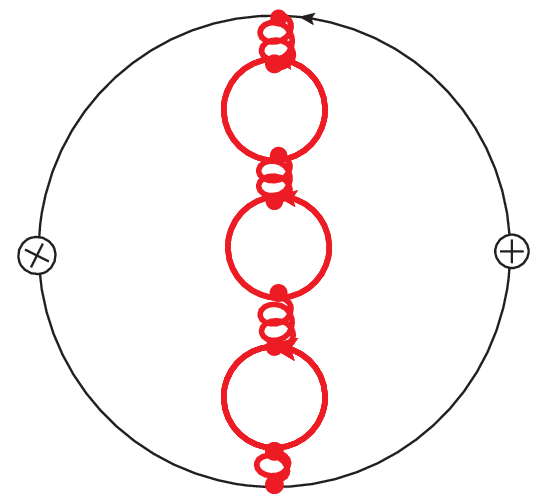}~~~~~~~~~~~~~ %
  \includegraphics[width=0.17\textwidth]{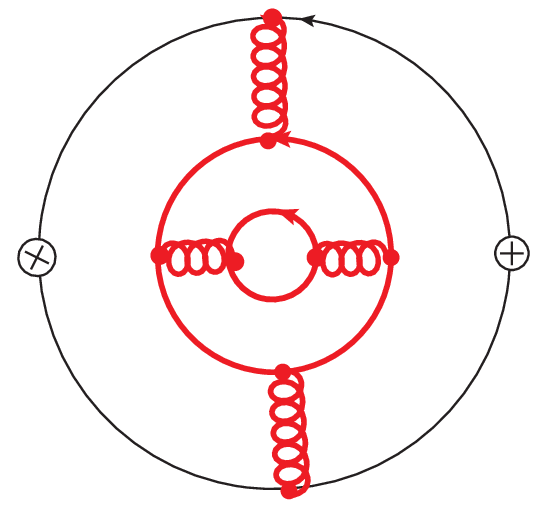}~~~~~~~~%
  \includegraphics[width=0.17\textwidth]{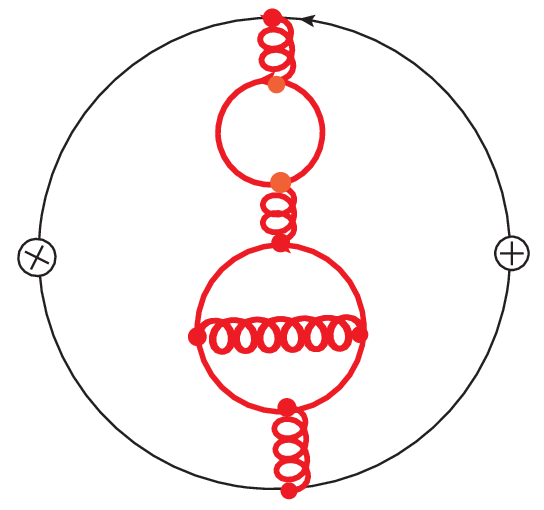}
 }
\vspace{-3mm} \caption{\footnotesize{\label{fig:fig2}
The types of the ``abelian'' diagrams for  $D_A$ in order $a_s^4$  of the self-energy type of only one gluon line (red in color) also include the symmetric diagram where the inner loops are interchanged.
\textbf{Left}: proportional to $C_\text{F} x^3$, $\sim \beta_0^3$ \textbf{Right}: proportional to $C_\text{F}^2 x^2$: $\Rightarrow \beta_2, \beta_1 \beta_0$} }
\end{figure}

Here we discuss  independent tests to verify
 the elements of $\{\beta \}$-expansion for $d_4 (c_4)$ obtained in \cite{Baikov:2022zvq} on
the basis of the results presented in \cite{Chetyrkin:2022fqk} within QCDe.
These results for the $d_4[.]$ elements are presented below,

\begin{subequations}
 \label{eq:d_4expr}
 \begin{eqnarray}
 d_4[3] &=& C_\text{F}\left(\frac{6131}9 - 406 \zeta_3 - 180 \zeta_5\right); \label{eq:d_43}\\
  d_4[1,1]&=& C_\text{F}\left(385 - \frac{1940}3 \zeta_3+ 144 \zeta_3^2 + 220 \zeta_5\right);\label{eq:d_411} \\
 d_4[0,0,1]&=& C_\text{F} \left(\frac{355}{6} + 136 \zeta_3 - 240 \zeta_5\right); \label{eq:d_4001} \\
 d_4[2] &=&-C_\text{F} \left[ C_\text{F} \left(\frac{6733}{8} + 1920 \zeta_3 - 3000 \zeta_5\right) + \right.\nonumber
 \\
         &&\left. \phantom{-C_\text{F}\Big[ }C_\text{A} \left(\frac{20929}{144} - \frac{12151}{6} \zeta_3 + 792
         \zeta_3^2
         + 1050 \zeta_5\right)\right]; \label{eq:d_42}
 \end{eqnarray}
 \begin{eqnarray}
 d_4[1] &=& C_\text{F}\bigg[
 -C_\text{F}^2 \left(\frac{447}2 - 42 \zeta_3 - 4920 \zeta_5 + 5040 \zeta_7\right)+\nonumber\\
 &&\phantom{C_\text{F} \bigg[ } C_\text{A} C_\text{F} \left(\frac{3301}4 - 678 \zeta_3 - 2280 \zeta_5 + 2520
 \zeta_7\right)+\nonumber\\
 &&\phantom{C_\text{F} \bigg[ }C_\text{A}^2 \left(\frac{16373}{36} - \frac{17513}{3} \zeta_3 + 2592 \zeta_3^2 +
    3030 \zeta_5 - 420 \zeta_7\right)\bigg] \label{eq:d_41}; \\
  d_4[0,1]&=& - C_\text{F} \left[C_\text{A} \left(\frac{139}{12} + \frac{1054}3 \zeta_3 - 460 \zeta_5\right) +
    C_\text{F} \left(\frac{251}4 + 144 \zeta_3 - 240 \zeta_5 \right)\right]; \label{eq:d_401}
 \end{eqnarray}
 \begin{eqnarray}
 d_4[0] &=&\tilde{d}_4[0]+ \delta d_4 \nonumber \\
         &=&C_\text{F}^4 \left(\frac{4157}{8} + 96 \zeta_3\right) -
 C_\text{A} C_\text{F}^3 \left(\frac{2409}{2} + 432 \zeta_3\right)+
 C_\text{A}^2 C_\text{F}^2 \left(\frac{3105}{4} + 648 \zeta_3\right) + \nonumber \\
 &&C_\text{A}^3 C_\text{F} \left(\frac{68047}{48} + \frac{8113}{2} \zeta_3 - 7110 \zeta_5\right) + \delta d_4\,;
 \label{eq:d_40}
 \end{eqnarray}
 \begin{eqnarray}
 \delta d_4&=&-16\Big[n_f \frac{d_F^{a b c d}d_F^{a b c d}}{d_F} \left(13 + 16 \zeta_3 - 40 \zeta_5\right) +
 \frac{d_A^{a b c d}d_F^{a b c d}}{d_F} \left(-3 + 4 \zeta_3 + 20\zeta_5 \right)\Big]\,. \label{eq:d_40delta}
\end{eqnarray}
 \end{subequations}
Equation(\ref{eq:d_43}) for $d_4[3]$  can also be independently confirmed by the partial result for $\beta_0^3 d_4[3]$ in Eq.(C.4) in \cite{Ball:1995ni}, see Fig.C1(V1) there, which agrees with the results in \cite{Broadhurst:1992si,Broadhurst:1993ru} too.
Here we reproduce these ``abelian'' diagrams in Fig.\ref{fig:fig2}(see left panel)  for reader's convenience.

(i)The projection of the weighted sum $ \beta_2\, d_4[0,0,1]+\beta_1 \beta_0\, d_4[1,1]$  onto the Casimir $C_F^2 x^{2}$ is
denote as $s_4=\hat{P}_{C^2_F x^2}\left( \beta_2\, d_4[0,0,1]+\beta_1 \beta_0\, d_4[1,1] \right)$, where $\hat{P}_{C^2_F x^2}$ means
the corresponding projector.
The evident diagrammatic origins just of these terms in $s_4$ are presented in Fig.C1(V2,V3) \cite{Ball:1995ni} that is
reproduced in Fig.\ref{fig:fig2}(Right) here, the corresponding partial results are presented in Eqs.(C5, C6) \cite{Ball:1995ni}.
The results for each of these two diagrams, evidently related to the factors $\beta_2,~ \beta_1 \beta_0$ due to gluon line
renormalizations, contributes to both the terms in the sum $s_4$.
The sum of the partial results of the direct QCD calculation of these diagrams  reads in the lhs\footnote{
up to the factor $4^{-4}$ in the lhs due to another definition of $a_s$, $a_s=\alpha_s/\pi$ in \cite{Ball:1995ni}} of the equation as
\ba \label{eq:part_cf2_x2}
\!\!\!\!\!\! C_F^2 x^2 \left(\frac{63250}{27} - 2784 \zeta_3 + 768 \zeta_3^2 \right)=s_4&\!\equiv\!&\! \hat{P}_{C^2_F x^2}
  \Big(\!\beta_2\,d_4[0,0,1]+\beta_1 \beta_0\, d_4[1,1]\! \Big) \\
                        &\!=\!&\!\left(\!\frac{44}{9}C_F x^2\!\!\right)\!d_4[0,0,1] + (-4C_F x)\!\left(\!-\frac{4}{3}x\!\right)\! d_4[1,1]\,. \nonumber
  \ea
  The abelian projections of the $\beta$-coefficients in the last equation correspond to the underlined terms in Eq.(\ref{eq:beta-coeff})  (Appendix \ref{App:A}).
Substituting the corresponding elements $d_4[.]$ from Eqs.(\ref{eq:d_411},\ref{eq:d_4001}) into the rhs, one obtains the equality with the lhs;
at the same time, the terms with $\zeta_5$ that exist in each of the elements $d_4[0,0,1], d_4[1,1]$ cancel in the sum $s_4$.
So the sum of these elements \textit{is confirmed}.
Recall that $\beta_2\, d_4[0,0,1]+\beta_1 \beta_0\,d_4[1,1]$ corresponds to special kinds of  dressing of only one gluon line in Fig. \ref{fig:fig2},
 % C1(V2,V3), \cite{Ball:1995ni},
 but it does not exhaust all the contributions to $C_F^2 x^{2}$.

(ii) To obtain the completed projection on $C_F^2 x^2$, presented in the lhs of (\ref{eq:cf2_x2}), see \cite{Baikov:2010je},
\be \label{eq:cf2_x2}
\!\!\!\!\!\!\!\! C_F^2 x^2 4\!\left(\!\frac{5713}{27} - \frac{4648}3 \zeta_3 + 192 \zeta_3^2 +\frac{4000}3\zeta_5\!\! \right)\!=\!\hat{P}_{C^2_F x^2}\!\Big(\!\! \beta_2 d_4[0,0,1]+\beta_1 \beta_0 d_4[1,1]+\beta_0^2 d_4[2]\!\Big),
 \ee
one should supplement $s_4$ with the contribution $\hat{P}_{C^2_F x^2}\left(\beta_0^2 d_4[2] \right)$ in the rhs.
The diagrams forming this additional term are not of ``the only one dressed gluon line''  type.
Equality (\ref{eq:cf2_x2}) unambiguously determines  the abelian part of the element $d_4[2]$ and confirms  now Eq.(\ref{eq:d_42}).
Concluding we confirm the values of the elements $d_4[0,0,1], d_4[1,1]$ and the abelian part of $d_4[2]$ based on the direct QCD calculations
\footnote{The corresponding values of $d_4[0,0,1],d_4[1,1]$ presented in \cite{Cvetic:2016rot}, and \cite{Kataev:2022iqf}, \cite{Goriachuk:2021ayq} (Table 2 there) do not agree with the lhs in (\ref{eq:part_cf2_x2}), the abelian part of their $d_4[2]$  does not agree with (\ref{eq:cf2_x2}) with this $s_4$.}.
%---------------------------------------------test--------------------

(iii) In the next step we will consider the contribution to $C_F\left(C_F^2 x\right)$ that is known from \cite{Baikov:2010je},
see the lhs of (\ref{eq:cf3_x}),
 \begin{subequations}
\ba
\!\!\!\!\! C_F^3 x\! \left(\!\frac{2002}{3} + 792 \zeta_3 - 8000\zeta_5\!+\! 6720 \zeta_7\right)\!&\!=\!&\!\hat{P}_{C_F^3 x}\!\Big( \beta_2\, d_4[0,0,1]+\beta_1\, d_4[0,1] + \beta_0\, d_4[1]\Big) \label{eq:cf3_x}\\
\!&\!=\!&\!\hat{P}_{C_F^3 x}\!\Big[\left(\!2 C_F^2 x\!\right) d_4[0,0,1]+\!\left(\!-4 C_F x\!\right) d_4[0,1] +\!\left(\!\!-\frac{4}3 x\!\right)\!d_4[1]\!\Big], \nonumber
 \ea
 while its rhs is  fulfilled by the weighted sum.
% $\beta_2 d_4[0,0,1]\!+\!\,\beta_1 d_4[0,1]\!+\!\,\beta_0 d_4[1]$
Of course, the values of elements $d_4[.]$ in Eq.(\ref{eq:d_4expr}) satisfy this equality.
Assuming that the value $d_4[0,0,1]$ is already confirmed from the condition (\ref{eq:part_cf2_x2}) in item (i),
%(the values of the elements in (\ref{eq:d_4expr}) satisfy that condition)
one obtains an independent test for the partial sum $\hat{P}_{C_F^3 x}\!\left(\beta_1 d_4[0,1]+\,\beta_0 d_4[1]\right)$ of the remaining elements,
\ba
 C_F^3 x\! \left(549 + 520 \zeta_3 - 7520\zeta_5+6720\zeta_7 \right)\!&\!=\!&\!-\hat{P}_{C_F^3 x}\!\Big[\left(4 C_F x \right) d_4[0,1] +\!\left(\frac{4}3 x\right) d_4[1]\Big]\,. \label{eq:cf3_xa}
 \ea
 \end{subequations}
To complete the confirmation for the  5-loop $d_4[.]$ elements, we need at least 3 independently calculated elements,
e.g., $d_4[0,0,1],\, d_4[1,1]$ and, say, $d_4[0,1]$.
In other words, taking into consideration the Casimir content of these elements, this parameterization consist of 4 independent scalar parameters\footnote{a similar parameterization was considered by A.Kataev, see A. Kataev's talk at the session of RAS\&JINR on 03.04.2024, https://indico.jinr.ru/event/4174/contributions/25722/}.
So to obtain the important element $d_4[0] (c_4[0])$, one should take $d_4(x_0)$ or $c_4(x_0)$, all the powers of $\beta_0$ cancel in the corresponding $\{\beta \}$-expansion,
see Eq.(\ref{eq:d_4},\ref{eq:x_0}), and one arrives at
\begin{subequations}
\label{eq:2.5}
\ba
d_4[0]&=&d_4(x_0)-\beta_2(x_0)\,d_4[0,0,1]-\beta_1(x_0)\,d_4[0,1]\,, \label{eq:2.5a}\\
c_4[0]&=&c_4(x_0)-\beta_2(x_0)\,c_4[0,0,1]-\beta_1(x_0)\,c_4[0,1]\, \label{eq:2.5b}\,.
\ea
 It is important  if the elements $d_4[0,0,1], d_4[0,1]$ are already known, then the counter part elements $c_4[0,0,1], c_4[0,1]$ are also known from the corollaries \cite{Baikov:2022zvq} of the CBK relation  \cite{Baikov:2022zvq,Chetyrkin:2022fqk}  like Eq.(\ref{eq:CBK1}).
%---------------------------------------------test---------------------
 \end{subequations}

 Let us mention here a rather rough test for  elements $d_{3,4}[0]$ that is related to a special  choice of Casimirs $C_A, C_F$, see (\ref{eq:A3}).
The coefficients of perturbative expansion based on $SU_c(N)$ dynamics are the polynomials over the  $C_A, C_F$,
let us consider these Casimirs as variables, which are related by the condition $7C_A+11C_F=0$ (that formally corresponds to the value $N=\sqrt{11}/5$).
 This linear condition, see Eqs.(\ref{eq:x_0}, \ref{eq:x_1}),
 leads to the evident nullification of the coefficients $\beta_0, \beta_1$,
\ba
\beta_0(x_0)=\beta_1(x_0) = \beta_0(x_1)=\beta_1(x_1)=0 \,. \label{eq:beta-add}
\ea
This condition  radically simplifies the color structure of elements and in virtue of (\ref{eq:beta-add}) leads
\ba
\!\!\!\!\!\text{at}~7 C_A+11 C_F=0:&& \nonumber \\
\!\!\!\!\!d_3(x_0)\!=\! d_3[0]&\!=\!& C_A^3\! \left(22607 + 313632 \zeta_3 \right) \frac{7}{15972}; \label{eq:d3add} \\
\!\!\!\!\!d_4(x_0)\!=\! d_4[0]+\beta_2(x_0)d_4[0,0,1]&\!=\!&C_A^4\!\left(\!\frac{308178983}{351384} + \frac{7782229}{29282} \zeta_3
  + \frac{22680}{121} \zeta_5\!\right)\!+\!\delta d_4.  \label{eq:d4add}
\ea
The Eqs.(\ref{D-30}) for $d_3[0]$, and (\ref{eq:d_4001}) for $d_4[0,0,1]$, (\ref{eq:d_40}) for $d_4[0]$ as well as the corresponding results  in \cite{Goriachuk:2021ayq}
satisfy  both the tests in Eqs.(\ref{eq:d3add},\ref{eq:d4add}).

% \newpage
 \section{ Reduction of $\{\beta\}$-expansion to a generalized NNA}
 \label{sec:4}
Here we consider a conjecture about the hierarchy of contributions to the perturbation coefficients $d_n,\, c_n$  based on the $\{\beta \}$--expansion, which generalizes the well-known NNA \cite{Broadhurst:1994se,Beneke:1994qe}.
 It  also generalizes  the results of the $\beta_0^n$-expansion  \cite{Lovett-Turner:1995zwc} and the $(\beta_0^n, \beta_1\beta_0^{n-2})$
 expansion in \cite{Ball:1995ni}.
This hierarchy rearranges the elements into a group that can  provide the main contribution to the coefficients of the expansion
$d_n (c_n)$.
It was first introduced in \cite{Mikhailov:2004iq} (Sec.4) and was discussed in
\cite{Bakulev:2010gm,Kataev:2014jba}.
We will discuss the relations between the main term of this hierarchy, the topological structure of the  diagrams corresponding to it
in the abelian limit,
and the conditions on the indices of the involved elements of the $\{\beta \}$-expansion.
  \subsection{Generalized NNA, illustration of $\beta_0$-dominance}
Here we will use the ordering of contributions in powers of $\beta_0$ under the condition of ``large $\beta_0$''
(in QCD $\beta_0 \sim 10$).
More precisely, the common factor $\beta_0^{n-1}$ in any order $n$ will be taken out of the expansion coefficient $d_n$ %brackets
to order all the elements over the inverse powers of $1/(\beta_0)^i$ \cite{Mikhailov:2004iq}.                     % within the brackets.
Recall that the well-known BLM \cite{Brodsky:1982gc} trick that was a predecessor of the $\{\beta \}$-expansion,
assumes for its  efficiency  the condition $\mid\beta_0 d_2[1]\mid \gtrsim  d_2[0]$, see the corresponding ratios in Eq.(\ref{eq:d_2numer}).

Then, we follow  the ``practical'' observation $\beta_i = O(\beta_0^{i+1})$ that works at $n_f=0\div 5$, $i=1\div 4$,
and invent the notation $\Ds b_i\stackrel{def}{=} \beta_i/\beta_0^{i+1}= O(1)$, see  (\ref{eq:num-b}).
Of course, the last equality should be broken down at higher orders $i$ of pQCD due to the further factorial growth of $\beta_i$.
So joining both these conditions, we  take into account the terms with the coefficients $b_i$ together with others in the same order of $1/(\beta_0)^i$, see the first brackets in Eqs.(\ref{eq:d_3numer}) and the first and the second brackets in Eqs.(\ref{eq:d_4numer}) below.
We call this kind of rearrangement in the groups of the same $1/\beta_0$-order in brackets  the generalized NNA (gNNA).
The expansion coefficients of the $D_A$-function and BpSR $S^{Bjp}$ are presented below as  the sums of these groups  accompanied by their numerical estimates in inscriptions (at $n_f=3$, NNA estimates are shown from below in red), \vspace*{-3mm}
\ba \label{eq:d_2numer}
d_2=\beta_0\Big[\underbrace{ d_2[1]}_{\RedTn{\bm{2.77}}}+\overbrace{\frac{1}{\beta_0}d_2[0]}^{\bm{0.15}}\Big]\approx \beta_0[\bm{2.92}];~
c_2=\beta_0\Big[\underbrace{ c_2[1]}_{\RedTn{\bm{-8}}}+\overbrace{\frac{1}{\beta_0}c_2[0]}^{\bm{1.63}}\Big]\approx
\beta_0 [\bm{-6.4}]
\ea
\begin{subequations}
\label{eq:d_3numer}
\ba
d_3&=&\beta_0^2\,\bigg[\overbrace{\Big(\underbrace{ d_3[2]}_{\RedTn{\bm{12.4}}}
     + b_1\,d_3[0,1]\Big)}^{\bm{8.6}}+ \overbrace{\frac{1}{\beta_0}d_3[1]}^{32.6}+\overbrace{\frac{1}{\beta_0^2}d_3[0]}^{-28.3}\bigg]\approx \beta_0^2\,\left[\,\bm{12.9}\, \right]\\
c_3&=&\beta_0^2\,\bigg[\overbrace{\Big(\underbrace{ c_3[2]}_{\RedTn{\bm{-25.6}}}
     + b_1\,c_3[0,1]\Big)}^{\bm{-25.9}}+ \overbrace{\frac{1}{\beta_0}c_3[1]}^{-17.8}+\overbrace{\frac{1}{\beta_0^2}c_3[0]}^{27.7}\bigg]\approx \beta_0^2\,\left[\,\bm{-16}\, \right]
\ea
\end{subequations}
\begin{subequations}
 \label{eq:d_4numer}
\ba
\!\!\!\!\!\!d_4
   &=\!&\!\beta_0^3\!\bigg[\overbrace{\Big(\underbrace{ d_4[3]}_{\RedTn{\bm{8.7}}}
     + b_1\,d_4[1,1]+ b_2\, d_4[0,0,1]
\Big)}^{\bm{24}}
     + \overbrace{\frac{1}{\beta_0}(d_4[2] + b_1  d_4[0,1])}^{34.3}
     +\overbrace{\frac{1}{\beta_0^2}d_4[1]}^{-6.8}
  +\overbrace{\frac{1}{\beta_0^3}d_4[0]}^{-34.35}\bigg]
\nonumber \\
   & \approx& \beta_0^3 \left[\, \bm{17.2}\, \right] \label{eq:d_4frag} \\
   & &  \nonumber  \\
\!\!\!\!\!\!c_4
   &=\!&\! \beta_0^3\!\bigg[\overbrace{\Big( \underbrace{c_4[3]}_{\RedTn{\bm{-89.6}}}
         + b_1\,c_4[1,1] + b_2\, c_4[0,0,1]\Big)}^{\bm{-112}}
     + \overbrace{\frac{1}{\beta_0}(c_4[2] + b_1  c_4[0,1])}^{-95.5}
     +\overbrace{\frac{1}{\beta_0^2}c_4[1]}^{136.5}
  +\overbrace{\frac{1}{\beta_0^3}c_4[0]}^{9.4}\bigg]\nonumber \\
  & \approx & \beta_0^3 \left[\,\bm{-61.7}\, \right]
       \label{eq:c_4frag}
       \ea
  \end{subequations}
As it is seen, the leading  $(1/\beta_0)^0$-groups numerically dominate in all of the sums in (\ref{eq:d_3numer},\ref{eq:d_4numer}),
  while the subleading groups for the most part mutually  cancel one another, compare the lhs and the rhs there.
   The evident manifistation of gNNA vs NNA can be seen in the leading group of (\ref{eq:d_4frag}).
The numerical estimates of each of the elements $d_{2,3,4}[.],~c_{2,3,4}[.]$ are shown in Appendix \ref{App:B},
where the diagrammatically confirmed results are marked in red.
 In addition, just the result of this leading group has been independently confirmed here in Sec.\ref{sec:sec3},
see the  rhs of Eq.(\ref{eq:part_cf2_x2}), which was
based on the calculations of the QCD diagrams  presented in Fig.\ref{fig:fig2}.
Therefore, the correspondence of gNNA and the total sum in (\ref{eq:d_4frag}) does not even depend on the detailed results in
Eqs.(\ref{eq:d_411},\ref{eq:d_4001}) obtained with the help of QCDe.
On the other hand, the contributions of the  suppressed in $1/\beta_0$ the so-called ``scale invariant'' terms $d_{3,4}[0]$
almost cancel with the other subleading groups.
At the same time,
these $d_i[0]~(c_i[0])$ do not correspond by themselves to the final sum in the rhs neither in value nor in sign
\footnote{The authors of PMC, see \cite{DiGiustino:2023jiq} and references therein, propagate another,
important role of the ``scale invariant'' elements $d_n[0]$ within the PMC approach.}.
 We recognize that our conjecture is based  on observations of a few examples only,
and the reason of the aforementioned mutual cancellations is not clear;
nevertheless, it looks reasonable to formulate the conjecture for the next orders of pQCD.

\subsection{Conjecture of $\beta_0$-dominance}
In the case of 6 loop \cite{Baikov:2022zvq}, the leading gNNA group is presented in the first line in Eq.(\ref{eq:d_5frag}) below.
Following our conjecture, one needs to calculate
\textit{restricted types} of the corresponding diagrams \textit{to estimate} the complete result $d_5$ based on (\ref{eq:d_5frag}),
\begin{subequations}
\begin{eqnarray}
d_5&=&\! \beta_0^4\!\bigg[\Big(\overbrace{d_5[4]+b_1 d_5[2,1]+ b_1^2\,d_5[0,2]+ b_2 d_5[1,0,1]+ b_3 d_5[0,0,0,1]
}^{p(5-1)}\Big)+ \label{eq:d_5frag}\\
  & &\phantom{\! \beta_0^4\!\bigg[\Big(}\frac{1}{\beta_0} \Big(d_5[3]+
     b_1 d_5[1,1]
     +  b_2 d_5[0,0,1]\Big) + \frac{1}{\beta_0^2} \Big( d_5[2]
     + b_1 d_5[0,1]\Big) + \label{eq:d_5frag-b} \\ %\nonumber
  & &\phantom{\! \beta_0^4\!\bigg[\Big(}\frac{1}{\beta_0^3}d_5[1]+\frac{1}{\beta_0^4}d_5[0]\bigg]\,.  \label{eq:d_5frag-c}
\ea
 \end{subequations}
 The number of elements of the leading group in Eq.(\ref{eq:d_5frag}) is equal to the number of partitions $p$ depending on perturbation order $n$, \eg, here -- $p(5-1)=\bm{5}$ \cite{OEIS}, see the detailed discussion of this correspondence in \cite{Baikov:2022zvq}, Sec.4.
  So taking a natural series for the argument of the function $p$, one
 obtains $p(1,2,3,\bm{4},5,6,7,\ldots)=$ $1,2,3,\bm{5},7,11,15,\ldots$
For the order $n$ of pQCD the leading gNNA group consists of the number of partitions $p(n-1)$ elements that can be traced from the dressing of \textit{only  one gluon line of the diagram}.
The indices of the elements $d_n[j_1,\ldots,j_{n-1}]$ of the leading group satisfy the evident equality
$1j_1+2j_2+\ldots\! +\! (\!n\!-1)\! j_{n-1}=n-1$,
while for the different subleading groups the conditions for the indices are  $1\, j_1+2\, j_2+\ldots + (n-1)\, j_{n-1} = n-2, n-3, \ldots$, where the shift from the number $n-1$ in the rhs determines the order of $1/\beta_0$ suppression
see, \eg, the groups in (\ref{eq:d_5frag-b}, \ref{eq:d_5frag-c}).

\begin{figure}[ht]
\centerline{\!\!\includegraphics[width=0.17\textwidth]{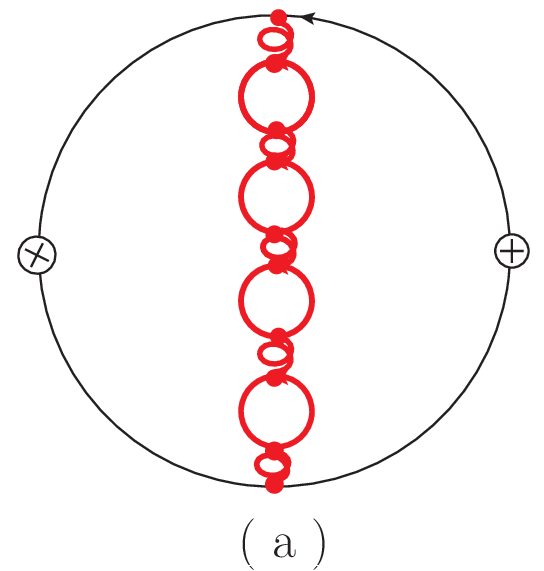}\!
  \includegraphics[width=0.17\textwidth]{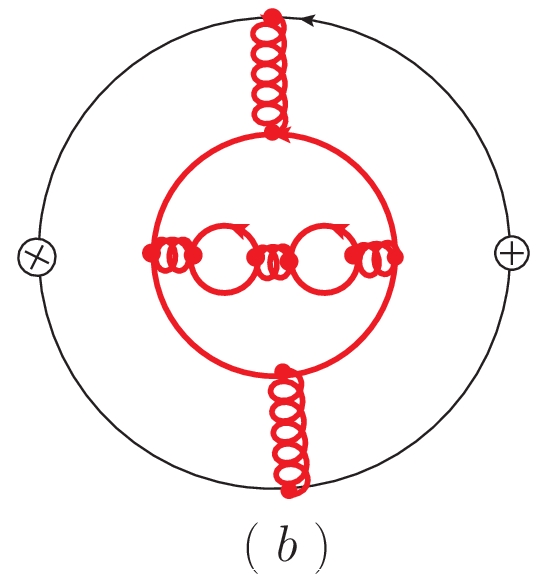}\!
  \includegraphics[width=0.17\textwidth]{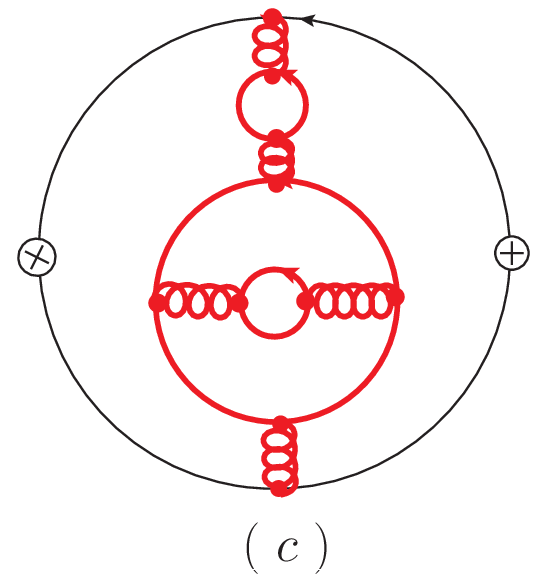}\!
 \includegraphics[width=0.17\textwidth]{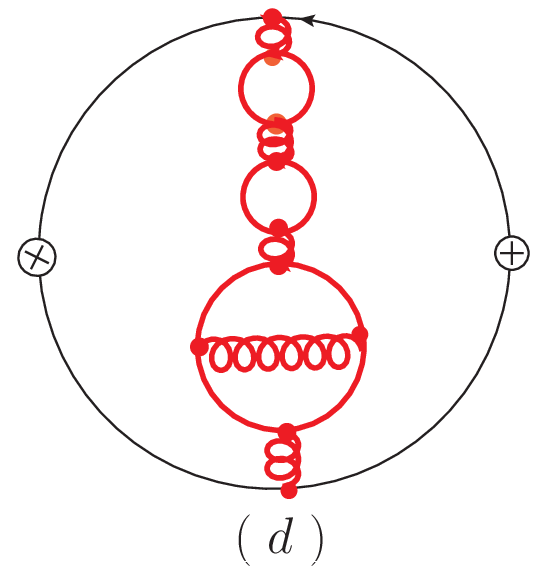}\!
 \includegraphics[width=0.17\textwidth]{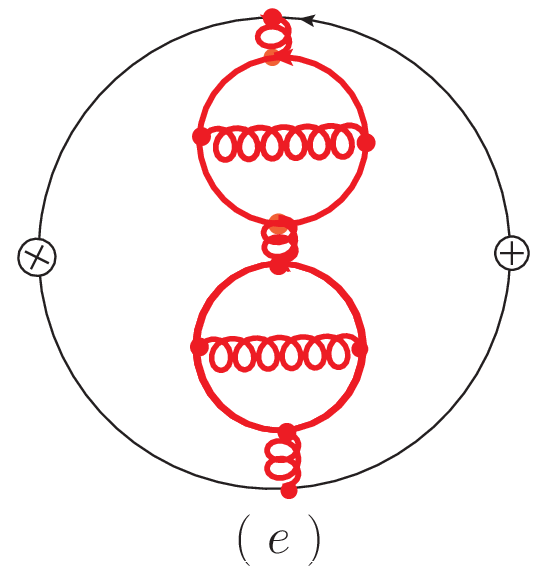}\!
 \includegraphics[width=0.17\textwidth]{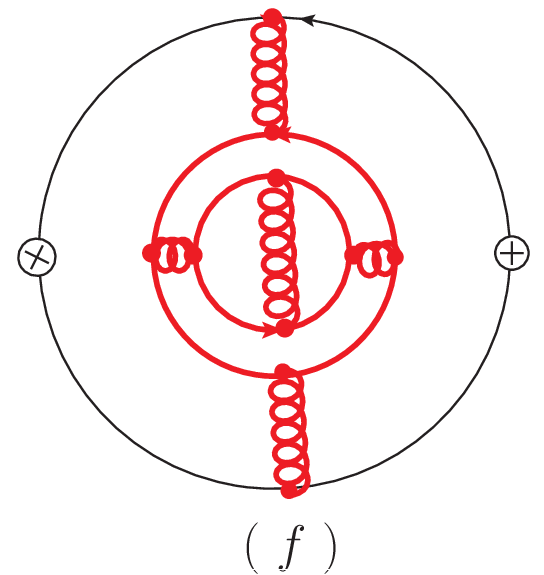}
 }
\vspace{-3mm} \caption{\footnotesize{\label{fig:fig3}
 The types of ``abelian'' diagrams corresponding to gNNA in order $a_s^5$ with all possible interchanges of the inner loops.
 (a) $C_\text{F}\, x^4$ $\sim \beta_0^4$; $C_\text{F}\, x^3$: (b), (c), (d) $\Rightarrow \beta_3,\, \beta_0\beta_2,\, \beta_0^2\beta_1$;\, $C_\text{F}^2\,x^2$: (e),(f) $\Rightarrow \beta_1 \beta_1, \beta_3$.}}
\end{figure}

If the  elements of the gNNA group in (\ref{eq:d_5frag}) are already obtained,
 they can be easily checked following the topological types of  ``abelian'' diagrams in Fig.\ref{fig:fig3},
%\noindent
where the chain of $n$ bubbles in Fig.\ref{fig:fig3}(a) definitely  corresponds to only $d_n[n-1]$.
The other diagrams -- $(b, c, d)$ contribute to different
combinations of the $\beta$-coefficients admissible for the leading group  that has a projection on $C_\text{F}\cdot x^3$,
while the diagrams $(e, f) \sim C_\text{F}^2\cdot x^2$. So one has
\begin{eqnarray}
\text{sum of diag.(b,c,d)}&=&P_{C_\text{F} x^3}\left(\beta_3 d_5[0,0,0,1]+\beta_2\beta_0\,d_5[1,0,1]+\beta_1\beta_0^2\,d_5[2,1] \right);\\
\text{sum of diag.(e,f)}&=&P_{C_\text{F}^2 x^2}\left(\beta_3 d_5[0,0,0,1]+\beta_1^2\,d_5[0,2]\right)\,.
\end{eqnarray}
%Recall, the contributions to (\ref{eq:d_5frag}) can be restored from the abelian projection of these diagrams results.
It is clear that the diagrams similar those in Figs.\ref{fig:fig2},\,\ref{fig:fig3} for the leading gNNA group can also be easily identified for the twist 2 coefficient function $C_{Bjp}$ of the observable  $S^{Bjp}$.

To conclude, we present  the leading gNNA  for any order $n\geqslant 4$ as a weighted  sum of $p(n-1)$ elements,
\begin{eqnarray}
d_{n}
\!\! &=&\!\!\beta_0^{n-1}
%   \underbrace{
\bigg[\Big(\!\overbrace{ d_{n}[n\!-\!1]+b_1d_{n}[n\!-\!3,1]+\!\ldots\!+b_{(n-2)}d_{n}[\underbrace{0,\ldots,1\!}_{n-1}]}^{p(n-1)}\! \Big)+\ldots + \frac{d_n[0]}{\beta_0^{n-1}} \bigg]\,.
%   }_{N(n)}\,.
\label{eq:d_n}
\end{eqnarray}
This group corresponds to the diagrams with  only one dressed gluon line  like those presented in Fig.\ref{fig:fig3}.
On the contrary, the subleading gNNA groups correspond to the diagrams with more than one number of (dressed) gluon lines.
It should be noted that taking into account the first two terms in the leading group of (\ref{eq:d_n}) to improve the BLM was first proposed in \cite{Ball:1995ni}.

Using another language based on the hierarchy of $n_f$ orders, the application of gNNA  requires taking into account in addition to the leading powers of $n_f$, a subseries  of subleading orders  also  originating from different sources.
%The most complicated diagrams with one dressed gluon line for polarized operator $\Pi$ were considered in details in %\cite{Mikhailov:2018udp},
%our partial case in Eq.(\ref{eq:d_n}) corresponds to consideration in Sec.4.3, Eqs.(4.12, 4.16) there.

 \section{Conclusion}
We have confirmed our results for the elements of the $\{\beta \}$-expansion in \cite{Baikov:2022zvq,Mikhailov:2004iq}:
 in order of $O(a_s^3)$ -- completely; in order of $O(a_s^4)$ we confirmed two elements,
$d_4[0,0,1], d_4[1,1]$, and the abelian part of $d_4[2]$ from \cite{Baikov:2022zvq},
and for both the cases -- basing on the known calculation \cite{Ball:1995ni} within the standard QCD only.

We proposed a conjecture of $\beta_0$-dominance in $d_n$.
The weighted sum of such elements $d_n[j_1,j_2,\ldots,j_{n-1}]$ dominates,
whose indices satisfy the condition: $\!\!1j_1\!+\!2j_2\!+\ldots \!+\! (\!n\!-1)\!j_{n-1}= n-1$,
their contributions are not suppressed by powers of $1/\beta_0$.
This condition one to one corresponds to allowing for the topological class of diagrams with renormalization of only one gluon line.
The conjecture, in a lucky case, can significantly reduce the number of diagrams considered  and
make the laborious process of estimating physical quantities in high loops of calculations easier.
\acknowledgments
The author  is thankful to A.~L. Kataev for drawing my attention to ref. \cite{Ball:1995ni} and also for the constructive criticism of the paper, to  N. Volchanskiy for the fruitful discussions.
I am very grateful to K. Chetyrkin for the clarifying the important items of this consideration.

%\newpage
\begin{appendix}

\appendix
\section{ QCD $\beta$-function }
 \label{App:A}
%\textbf{1.}
The well-known renormalization group equation for the QCD coupling constant $\Ds a_s(\mu^2)\!=\!\frac{\alpha_s(\mu^2)}{4\pi}$ reads
\begin{equation}
\label{eq:beta}
\mu^2\frac{d a_s(\mu^2)}{d \mu^2}=
-\beta(a)=-a_s^{2}(\mu^2) \sum_{i\geq 0} \beta_{i}~ a_s^{i}(\mu^2)\,.
\end{equation}
The explicit expressions for the first coefficients of the $\beta$-function expansion with the underlined abelian terms read

\begin{subequations}
 \label{eq:beta-coeff}
\begin{eqnarray}
    \beta_0 &=& - \frac{4}{3}\,\underline{T_R n_f}+ \frac{11}{3}\,C_\text{A}
    \,;\qquad \label{eq:A1}\\
    \beta_1 &=& - 4\,\underline{C_\text{F} T_R n_f} + \frac{34}{3}\,C_{\text{A}}^{2}
              - \frac{20}{3}\,C_\text{A}T_R n_f ; \label{eq:A2} \\
    \beta_2 &=&\frac{44}{9} \underline{C_F (T_R n_f)^2}+2\, \underline{C_F^2 T_R n_f}  + \frac{2857}{54} C_A^3
 - \frac{205}{9} C_F C_A T_R n_f
 - \frac{1415}{27} C_A^2 T_R n_f
 \nonumber \\
 &&
 + \frac{158}{27} C_A (T_R n_f)^2\, ; \label{eq:A3}
\end{eqnarray}
\begin{eqnarray}
\beta_3 &=&\frac{1232}{243}\, \underline{C_F ( T_R n_f )^3} +\underline{( C_F T_R n_f )^2} \left( \frac{1352}{27} - \frac{704}{9} \zeta_3\right)+ 46 \underline{(C_F)^3 T_R n_f}+
C_A^3 T_R n_f \left( - \frac{39143}{81} + \frac{136}{3} \zeta_3 \right)
 \nonumber \\
 && +C_A C_F T_R^2 n_f^2 \left( \frac{17152}{243} + \frac{448}{9} \zeta_3 \right)
 + C_A C_F^2 T_R n_f \left( - \frac{4204}{27} + \frac{352}{9} \zeta_3 \right) + \frac{424}{243} C_A
 T_R^3 n_f^3 \nonumber \\
&&+ C_A^2 C_F T_R n_f \left( \frac{7073}{243} - \frac{656}{9} \zeta_3 \right)
+ C_A^2 T_R^2 n_f^2 \left( \frac{7930}{81} + \frac{224}{9} \zeta_3
\right) + C_A^4 \left( \frac{150653}{486} - \frac{44}{9} \zeta_3 \right)\nonumber \\
&& + n_f \dRANA \left( \frac{512}{9} - \frac{1664}{3} \zeta_3 \right) + n_f^2
\dRRNA \left( - \frac{704}{9} + \frac{512}{3} \zeta_3
\right) \nonumber \\
&&+ \dAANA \left( - \frac{80}{9} + \frac{704}{3} \zeta_3 \right),\label{eq:A4}
\end{eqnarray}
\end{subequations}
with the $SU_{c}(N)$-group fundamental fermion invariants

\begin{eqnarray}
\label{eq:inv}
&&T_R = \frac{1}{2}\, ; \; C_\text{F}= T_R\frac{N^2-1}{N}\, ; \; C_A =N\, ;\;
N_A = 2C_\text{F} C_\text{A} \equiv N^2-1\, ; \label{eq:A3} \\
&&d^{abc}d^{abc}= \frac{(N^2-4)}{N}N_A\, ; ~\dRA = \frac{N(N^2+6)}{48}N_A\, ; \nonumber \\
&&\dRR = \frac{N^4-6N^2+18}{96N^2}N_A\, ; ~ \dAA = \frac{N^2(N^2+36)}{24}N_A\, ,
\end{eqnarray}
$d_R$ is the dimension of the quark color representation,
$d_R = 3$ in QCD and $n_f$ denotes
the number of active flavors.
At $n_f=3$,
\be \label{eq:num-b}
\beta_0=9;~~b_1 \approx 0.79;~~b_2 \approx 0.88;~~b_3\approx 1.9.%~~b_4=\mbox{unknown},
\ee
\section{Numerical estimates of the elements $d_{2,3,4}[.],c_{2,3,4}[.]$ of $\{\beta \}$-expansion }
 \label{App:B}
We present the numerical estimates for the each of elements, discussed in Sec.\ref{sec:4} for reader's convenience.
For the expressions $d_{2,3}[.]$ in Eq.(\ref{D-21},\ref{eq:d1-4}) and similar ones $c_{2,3}[.]$ \cite{Baikov:2022zvq}
we have \cite{Kataev:2014zha,Kataev:2014zwa},
\begin{subequations}
 \label{eq:d3-numbers}
\begin{eqnarray}
\!\!\!\!\RedTn{d_2[1]}&\approx&\RedTn{\bm{2.77}},~~~~~ \RedTn{d_2[0]}=\RedTn{\bm{4/3}},~~~~~~~\RedTn{c_2[1]}=\RedTn{\bm{-8}},~~~~~~ \RedTn{c_2[0]}= \RedTn{\bm{44/3}}; \\
\!\!\!\!\RedTn{d_3[2]}&\approx&\RedTn{\bm{12.41}},~~ \RedTn{d_3[0,1]}\approx \RedTn{\bm{-4.799}}, d_3[1]\approx 293.55,~~ d_3[0]\approx -2295.84\,,
 \\
\!\!\!\!\RedTn{c_3[2]}&\approx&\RedTn{\bm{-25.56}}, ~\RedTn{c_3[0,1]\approx \bm{-0.43}}, c_3[1]\approx -159.84, ~c_3[0]\approx 2242.51\,.
\end{eqnarray}
 \end{subequations}
From the expressions for $d_4[.]$ in Eq.(\ref{eq:d_4expr}) one has
\begin{subequations}
 \label{eq:d4-numbers}
  \begin{eqnarray}
\RedTn{d_4[3]}&\approx&\RedTn{\bm{8.720}},~\RedTn{d_4[1,1]}\approx \RedTn{\bm{58.487}},~\RedTn{d_4[0,0,1]}\approx \RedTn{\bm{-34.955}}        ,\\
\RedTn{d_4[2]}&\approx&\RedTn{\bm{154.47}},~d_4[0,1]\approx 195.46,~d_4[1]\approx -550.54, \\
d_4[0]&\approx&-25105.61 + 20.5427\, n_f \approx -25043.97\,(\text{at}~ n_f=3)
\end{eqnarray}
 \end{subequations}
\begin{subequations}
 \label{eq:c4-numbers}
  \begin{eqnarray}
\RedTn{c_4[3]}&\approx&\RedTn{\bm{-89.63}},~c_4[1,1]\approx -61.647,~\RedTn{c_4[0,0,1]}\approx \RedTn{\bm{29.722}}        ,\\
c_4[2]&\approx&-707.165,~c_4[0,1]\approx -193.028,~c_4[1]\approx 11053.7, \\
c_4[0]&\approx&6871.02\,(\text{at}~ n_f=3)
\end{eqnarray}
 \end{subequations}
The elements marked in red in (\ref{eq:d3-numbers},\ref{eq:d4-numbers},\ref{eq:c4-numbers}) have been confirmed,
at least in their abelian part, by the diagrammatical calculations in the standard QCD in
Secs.\ref{sec:2.2}, \ref{sec:sec3}.
The widely debated ``zero'' element $d_4[0]$ reads $-98.069 + 0.0802\, n_f $ in the case of
the using the coupling constant $\alpha_s/\pi$ instead of $a_s$.

\end{appendix}
%\newpage
%\bibliographystyle{apsrev}
%\bibliographystyle{JHEP}
%\bibliography{multiloop_calc,conf-mod}
%\end{document}
%\begin{thebibliography}{20}
\newcommand{\noopsort}[1]{} \newcommand{\printfirst}[2]{#1}
  \newcommand{\singleletter}[1]{#1} \newcommand{\switchargs}[2]{#2#1}
\providecommand{\href}[2]{#2}\begingroup\raggedright

\end{document}